\documentclass[twoside,twocolumn,english,pra,twocolumn,aps,superscriptaddress]{revtex4-1}

\usepackage[T1]{fontenc}
\usepackage[latin9]{inputenc}
\usepackage{xcolor}
\usepackage{babel}
\usepackage{mathtools}
\usepackage{bm}
\usepackage{amsmath}
\usepackage{amsthm}
\usepackage{graphicx}
\usepackage[unicode=true,pdfusetitle,
 bookmarks=true,bookmarksnumbered=false,bookmarksopen=false,
 breaklinks=false,pdfborder={0 0 0},pdfborderstyle={},backref=false,colorlinks=true]
 {hyperref}

\makeatletter
\theoremstyle{plain}

\usepackage{tabularx}
\usepackage{MnSymbol}

\usepackage{makecell}
\usepackage{fontenc}

\hypersetup{urlcolor=blue}

\usepackage{cmap}
\usepackage[T1]{fontenc}

\input glyphtounicode
\usepackage{mathtools}
\usepackage{titlesec}

\makeatother

\begin{document}
\preprint{This line only printed with preprint option}
\title{Double-port measurements for robust quantum optical metrology}
\author{Wei Zhong}
\affiliation{Institute of Quantum Information and Technology, Nanjing University
of Posts and Telecommunications, Nanjing 210003, China}
\affiliation{National Laboratory of Solid State Microstructures, Nanjing University,
Nanjing 210093, China}
\author{Lan Zhou}
\affiliation{School of Science, Nanjing University of Posts and Telecommunications,
Nanjing 210003, China}
\author{Yu-Bo Sheng}
\email{shengyb@njupt.edu.cn}

\affiliation{Institute of Quantum Information and Technology, Nanjing University
of Posts and Telecommunications, Nanjing 210003, China}
\begin{abstract}
It has been proposed and demonstrated that path-entangled Fock states
(PEFSs) are robust against photon loss over NOON states\textcolor{teal}{{}
}{[}S. D. Huver \emph{et al.}, Phys. Rev. A \textbf{78}, 063828 (2008){]}.
However, the demonstration was based on a measurement scheme which
was yet to be implemented in experiments. In this work, we quantitatively
illustrate the advantage of PEFSs over NOON states in the presence
of photon losses by analytically calculating the quantum Fisher information.
To realize such an advantage in practice, we then investigate the
achievable sensitivities by employing three types of feasible measurements:
parity, photon-number-resolving, and homodyne measurements. We here
apply a double-port measurement strategy where the photons at each
output port of the interferometer are simultaneously detected with
the aforementioned types of measurements.
\end{abstract}
\maketitle

\section{Introduction}

An essential task in quantum optical metrology is to reach the ultimate
sensitivity limit to the phase measurement imposed by quantum theory
\citep{Lee2002JOP,Giovannetti2004,Giovannetti2006PRL,Giovannetti2011,Degen2017RMP,Pezze2018RMP,Liu2019JPA,Polino2020review}.
To date, extensive strategies have been proposed to improve the sensitivities
of phase measurements. Among these, one celebrated strategy is the
use of NOON states \citep{Lee2002JOP}
\begin{equation}
\vert N\!::\!0\rangle=\frac{1}{\sqrt{2}}\left(\vert N,0\rangle+\vert0,N\rangle\right).\label{eq:NOON}
\end{equation}
It has been experimentally demonstrated that this strategy can acquire
a Heisenberg-scaling sensitivity in lossless optical interferometry
\citep{Rarity1990PRL,Walther2004nature,Mitchell2004nature,Nagata2007science,Xiang2011NP},
which is a $\sqrt{N}$ factor improvement over the shot-noise limit
(SNL) \citep{Lee2002JOP,Giovannetti2004,Giovannetti2006PRL,Giovannetti2011,Degen2017RMP,Pezze2018RMP}.
However, such a quantum improvement will be completely gone even if
an individual photon loss occurs \citep{Dorner2009PRL,Escher2011NP,Demkowicz-Dobrzanski2012nc}.

To mitigate this problem, two main kinds of protocols have been raised:
active and passive. The former tries to actively reduce the effects
of losses by managing experimental processes, e.g., applying quantum
error correction \citep{Michael2016PRX,Zhou2018NC}. The latter tries
to passively retain sub-SNL sensitivities by using certain probe states
which are more robust against losses but at the expense of a bit of
sensitivity \citep{Huver2008PRA,Lee2009PRA,Dorner2009PRL,Kolodyfmmodecutenlseniski2010PRA,Demkowicz-Dobrzanski2009PRA,Knysh2011PRA,Jiang2012PRA}.
Among these passive protocols, one of the most representative proposed
by Huver \emph{et al.} is that of taking the path-entangled Fock states
(PEFS) \citep{Huver2008PRA}
\begin{equation}
\vert m\!::\!n\rangle=\frac{1}{\sqrt{2}}\left(\vert m,n\rangle+\vert n,m\rangle\right),\quad\left(m>n\right)\label{eq:MNNM}
\end{equation}
as the probe state. The authors in \citep{Huver2008PRA} demonstrated
that the PEFSs show more robustness against photon loss than NOON
states under the constraint of $\Delta\equiv m-n=N$ and referred
to it as robust optical metrology. However, the demonstration was
based on a measurement scheme which was yet to be implemented. Later,
Jiang \emph{et al.} considered the single-parity (SP) measurement
scheme, but failed to observe the robustness of the PEFS scheme compared
to the NOON state scheme \citep{Jiang2012PRA}. Whether or not the
PEFSs are robust to loss over NOON states in practice is still an
open question.

In this paper, we address this issue by finding effective measurements
to present the robustness of the PEFS scheme in practical metrological
experiments. To quantitatively demonstrate the advantage of the PEFS
scheme, we first derive the lower sensitivity bounds in the presence
of photon losses by invoking quantum Fisher information (QFI), since
the QFI is a figure of merit to indicate the performance of a metrological
scheme. We then discuss the achievable sensitivities with three different
types of measurements: parity, photon-number-resolving, and homodyne
measurements. All these measurements are achievable in today's state-of-the-art
experiments. Here we employ a double-port measurement protocol, in
which the photons at each output port of the interferometer are simultaneously
detected with the three aforementioned types of measurements. The
reason why we utilize this protocol comes from consideration of the
superiority of the double-port measurement over the single-port one
(see Appendix A for a demonstration of this superiority). Although
all aforementioned measurements were systematically studied in lossless
cases \citep{Pezze2008PRL,Hofmann2009PRA,Zhong2017PRA,Vidrighin2014NC,Rubio2019NJP},
there is little knowledge of the effect of these measurements in realistic
experiments where losses are present. Our work serves to complement
studies in this aspect. Thus it will provide valuable information
on the choice of measurement scheme which is more adequately implemented
in the laboratory.

This paper is organized as follows. In Sec. II we revisit the robust
optical metrology with PEFSs proposed by Huver \emph{et al.} \citep{Huver2008PRA}
and derive the lower sensitivity bounds based on quantum estimation
theory. In Sec. III we discuss the achievable sensitivities with three
different types of measurement. A summary and our conclusions are
given in Sec. IV.
\begin{figure}[b]
\centering{}\includegraphics[scale=0.28]{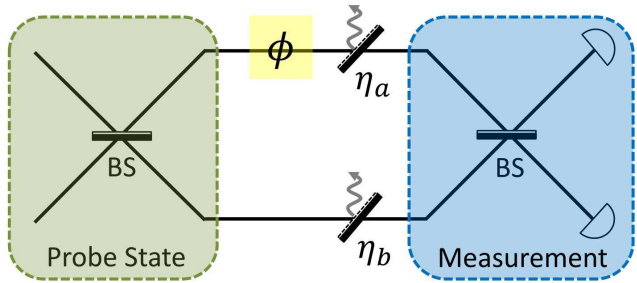}\caption{(Color online) Schematic of the optical two-arm interferometric setup
in the presence of photon losses. \label{fig:MZI}}
\end{figure}

\section{Robust quantum-optical interferometry with PEFSs}

A general optical two-mode interferometer consists of three optical
elements: two beam splitters (BSs) denoted by $\hat{B}$ and a phase
shifting denoted by $\hat{U}$ (see Fig.~\ref{fig:MZI}). This device
serves to illustrate the principle of phase estimation. The input
light is divided into two beams through the first BS. Then they accumulate
an unknown phase $\phi$ of interest under the phase shifting. The
value of the phase parameter is finally read out by detecting the
photons in the beams out from the second BS. 

From quantum estimation theory \citep{Helstrom1976Book,Holevo1982Book,Braunstein1994PRL},
the sensitivity of estimating the phase $\phi$ is statistically measured
by the unit-corrected mean-square deviation of the estimator $\phi_{{\rm est}}$
from the true value $\phi$,
\begin{equation}
\left(\delta\phi_{{\rm est}}\right)^{2}=\left\langle \left(\frac{\phi_{{\rm est}}}{\partial_{\phi}\langle\phi_{{\rm est}}\rangle_{{\rm av}}}-\phi\right)^{2}\right\rangle _{{\rm av}},
\end{equation}
where $\langle\cdot\rangle_{{\rm av}}$ denotes the statistical average
and the derivative $\partial_{\phi}\langle\phi_{{\rm est}}\rangle_{{\rm av}}$
removes the local difference in the ``units'' of $\phi_{{\rm est}}$
and $\phi$. Whichever measurement scheme is employed, the ultimate
limit to the sensitivity of the unbiased estimator is given by the
quantum Cram\'er-Rao bound
\begin{equation}
\left(\delta\phi_{{\rm est}}\right)^{2}\geq\frac{1}{\upsilon F_{Q}},\label{eq:QCRB}
\end{equation}
where $\upsilon$ is the repetition of the experiment and $F_{Q}$
is the so-called QFI \citep{Helstrom1976Book,Holevo1982Book,Braunstein1994PRL}.
This bound is asymptotically achieved for large $\upsilon$ under
optimal measurements, followed by the maximum likelihood estimator
\citep{Helstrom1976Book,Holevo1982Book,Braunstein1994PRL}. Although
optimal measurements for saturating this bound have been formally
demonstrated in Refs.~\citep{Braunstein1994PRL,Zhong2014JPA}, they
may be not achievable in realistic experiments. 

Assuming that $a$ and $b$ denote the annihilation operators of upper
and lower modes, respectively, we then define the Schwinger representation
\citep{Yurke1986PRA} as

\begin{eqnarray}
\hat{J}_{x}=\frac{1}{2}\left(a^{\dagger}b+ab^{\dagger}\right), & \quad & \hat{J}_{y}=\frac{1}{2i}\left(a^{\dagger}b-ab^{\dagger}\right),\\
\hat{J}_{z}=\frac{1}{2}\left(a^{\dagger}a-b^{\dagger}b\right),
\end{eqnarray}
which satisfy the commutation relations for the Lie algebra of $su\left(2\right)$,
\begin{eqnarray}
[\hat{J}_{x},\thinspace\hat{J}_{y}]=i\hat{J}_{z}, & \quad[\hat{J}_{y},\thinspace\hat{J}_{z}]=i\hat{J}_{x}, & \quad[\hat{J}_{z},\thinspace\hat{J}_{x}]=i\hat{J}_{y}.
\end{eqnarray}
and commute with the total photon number operator $\hat{N}=a^{\dagger}a+b^{\dagger}b$.
For simplicity, we define $\hat{J}_{0}=\hat{N}/2$ below. Within this
representation, the operations of the BS and phase shifting are represented
by
\begin{eqnarray}
\hat{B}=\exp\left(-i\frac{\pi}{2}\hat{J}_{y}\right), & \quad & \hat{U}=\exp\left(i\phi a^{\dagger}a\right),
\end{eqnarray}
where $\hat{B}$ refers to a balanced BS and $\hat{U}$ accounts for
the phase imprinting solely on mode $a$. An interferometry with asymmetric
BSs has been discussed recently for various states \citep{Preda2019PRA,Zhong2020}.
As for phase shifting $\hat{U}$, another both-arm configuration is
represented by $\exp\left(i\phi\hat{J}_{z}\right)$, which has been
widely considered in previous studies. Although these two types of
phase shifting have a subtle difference in phase estimation \citep{Jarzyna2012PRA},
they have the same effect in our cases. This is because the states
of consideration here are definite photon numbers and thus the phase
shifts accumulating by single- and both-arm phase shifting are up
to an irrelevant global phase factor, i.e., $\exp\left(i\phi N/2\right)$. 

In our work, as shown in Fig.~\ref{fig:MZI}, the probe state refers
to the state after the first BS, i.e., $\vert\psi\rangle=\hat{B}\vert\psi_{{\rm in}}\rangle$,
where $\vert\psi_{{\rm in}}\rangle$ is the input state powering at
the input port of the interferometer. After the phase shifting $\hat{U}$,
it becomes $\vert\psi\left(\phi\right)\rangle=\hat{U}\vert\psi\rangle$
acquiring an unknown phase to be estimated. Taking the NOON state
of Eq.~\eqref{eq:NOON} as the probe state, one can acquire the Heisenberg-limit
sensitivity associated with $F_{Q}=N^{2}$. However, in a practical
situation, the interferometry is often subjected to photon losses.
This may cause the dedicated NOON state to rapidly lose its quantum
advantage for phase resolution. To circumvent this problem, Huver
\emph{et al.} proposed a robust metrological scheme by applying the
PEFS as the probe state \citep{Huver2008PRA}. In the lossless case,
the QFI for the PEFS of Eq.~\eqref{eq:MNNM} reads $F_{Q}=\Delta^{2}$. 

Below, we revisit this robust metrological scheme and derive the sensitivity
bounds in the presence of photon losses by using the QFI. Here, photon
losses are modeled by inserting fictitious BSs with transmissivities
$\eta_{x}\:\left(x=a,b\right)$ into two arms of the interferometer
\citep{Huver2008PRA,Dorner2009PRL,Demkowicz-Dobrzanski2009PRA}. Their
effects can be formally represented in a Kraus form as
\begin{eqnarray}
\rho & = & \sum_{l_{a},l_{b}=0}^{\infty}\hat{K}_{a,l_{a}}\hat{K}_{b,l_{b}}\rho_{{\rm in}}\hat{K}_{b,l_{b}}^{\dagger}\hat{K}_{a,l_{a}}^{\dagger},
\end{eqnarray}
with Kraus operators 
\begin{eqnarray}
\hat{K}_{x,l} & = & \left(1-\eta_{x}\right)^{l/2}\left(\sqrt{\eta_{x}}\right)^{x^{\dagger}x}x^{l}/\sqrt{l!}.
\end{eqnarray}
Due to the commutation relationship between phase shifting and photon
losses \citep{Dorner2009PRL,Demkowicz-Dobrzanski2009PRA}, one can
freely exchange the order of operations between the phase shift and
the photon loss. For simplicity, we assume that photon losses act
after the phase shifting, as shown in Fig.~\ref{fig:MZI}. Successively
going through the phase shifting and the photon losses, the probe
state of Eq.~\eqref{eq:MNNM} of $\Xi\equiv m+n$ evolves into a
parametric mixed state as \citep{Huver2008PRA,Jiang2012PRA}

\begin{eqnarray}
\rho\left(\phi\right) & = & \frac{1}{2}\sum_{l_{a}=0}^{\Xi}\sum_{l_{b}=0}^{\Xi-l_{a}}\Big[B_{l_{a}l_{b}}^{m}C_{l_{a}l_{b}}^{m}\vert m-l_{a},n-l_{b}\rangle\langle m-l_{a},n-l_{b}\vert\nonumber \\
 &  & +B_{l_{a}l_{b}}^{n}C_{l_{a}l_{b}}^{n}\vert n-l_{a},m-l_{b}\rangle\langle n-l_{a},m-l_{b}\vert\nonumber \\
 &  & +\sqrt{B_{l_{a}l_{b}}^{m}B_{l_{a}l_{b}}^{n}}C_{l_{a}l_{b}}^{m}C_{l_{a}l_{b}}^{n}e^{i\Delta\varphi}\vert m-l_{a},n-l_{b}\rangle\nonumber \\
 &  & \times\langle n-l_{a},m-l_{b}\vert+{\rm H.c}.\Big],\label{eq:rho_phi}
\end{eqnarray}
with
\begin{eqnarray}
B_{l_{a}l_{b}}^{k} & \equiv & \binom{k}{l_{a}}\binom{\Xi-k}{l_{b}}\eta_{a}^{k}\left(\eta_{a}^{-1}-1\right)^{l_{a}}\eta_{b}^{\Xi-k}\left(\eta_{b}^{-1}-1\right)^{l_{b}},\quad\quad\\
C_{l_{a}l_{b}}^{k} & \equiv & H\left[k-l_{a}\right]-H\left[k-\Xi+l_{b}-1\right],
\end{eqnarray}
where $\binom{\bullet}{\bullet}$ denotes the binomial coefficient
and $H\left[n\right]$ is the Heaviside step function of a discrete
form. For NOON states, i.e., $m=N$ and $n=0$, Eq.~\eqref{eq:rho_phi}
can be simplified as a direct sum form of 
\begin{equation}
\rho\left(\phi\right)=\vert\xi\left(\phi\right)\rangle\langle\xi\left(\phi\right)\vert\oplus\rho_{D},\label{eq:lossy-NOON}
\end{equation}
where the $\phi$-dependent state $\vert\xi\left(\phi\right)\rangle$
is given by
\begin{eqnarray}
\vert\xi\left(\phi\right)\rangle & = & \frac{1}{\sqrt{2}}\left(\sqrt{\eta_{a}^{N}}e^{iN\varphi}\vert N,0\rangle+\sqrt{\eta_{b}^{N}}\vert0,N\rangle\right),\label{eq:NOON-sub}
\end{eqnarray}
up to a normalization constant and another part is a $\phi$-independent
diagonal matrix of dimension $2N$ as
\begin{eqnarray}
\rho_{D} & = & \frac{1}{2}\sum_{l=1}^{N}\big(B_{l0}^{N}\vert N-l,0\rangle\langle N-l,0\vert+B_{0l}^{0}\vert0,N-l\rangle\langle0,N-l\vert\big).\nonumber \\
\label{eq:NOON-subD}
\end{eqnarray}

To calculate the QFI, one should determine the eigenvalues and eigenstates
of $\rho\left(\phi\right)$ of Eq.~\eqref{eq:rho_phi}. However,
an analytical diagonalization of $\rho\left(\phi\right)$ may be not
easily obtained in the case with photon losses in both beams. Following
Ref.~\citep{Dorner2009PRL}, we obtain the upper bound of the QFI
for PEFSs, when losses occur in two arms, as

\begin{eqnarray}
F_{Q} & \leq & 2\!\left(m^{2}\!+\!n^{2}\right)-2\!\sum_{l_{a}=0}^{\Xi}\sum_{l_{b}=0}^{\Xi-l_{a}}\frac{\left(mB_{l_{a}l_{b}}^{m}C_{l_{a}l_{b}}^{m}+nB_{l_{a}l_{b}}^{n}C_{l_{a}l_{b}}^{n}\right)^{2}}{B_{l_{a}l_{b}}^{m}C_{l_{a}l_{b}}^{m}+B_{l_{a}l_{b}}^{n}C_{l_{a}l_{b}}^{n}}.\nonumber \\
\label{eq:QFI-PEFS-two}
\end{eqnarray}
This inequality is saturated when losses occur only in one arm, i.e.,
$\eta_{b}=1$, in which case, Eq.~\eqref{eq:QFI-PEFS-two} reduces
to \citep{Dorner2009PRL} 
\begin{eqnarray}
F_{Q} & = & 2\!\left(m^{2}\!+\!n^{2}\right)-2\!\sum_{l_{a}=0}^{\Xi}\!\!\frac{\left(mB_{l_{a}0}^{m}C_{l_{a}0}^{m}+nB_{l_{a}0}^{n}C_{l_{a}0}^{n}\right)^{2}}{B_{l_{a}0}^{m}C_{l_{a}0}^{m}+B_{l_{a}0}^{n}C_{l_{a}0}^{n}}.\quad\quad\label{eq:QFI-one}
\end{eqnarray}
In addition, it is also saturated for lossy NOON states of Eq.~\eqref{eq:lossy-NOON}
as 
\begin{eqnarray}
F_{Q}^{{\rm NOON}} & = & 2N^{2}\left(\frac{\eta_{a}^{N}\eta_{b}^{N}}{\eta_{a}^{N}+\eta_{b}^{N}}\right),\label{eq:QFI-NOON-two}
\end{eqnarray}
This can be easily obtained by employing the property that the QFI
for a density matrix of direct sum form $\rho\left(\phi\right)=\bigoplus_{i=1}^{n}\rho_{i}\left(\phi\right)$
is given by the sum over all amounts of QFI in terms of each submatrices,
i.e., $F_{Q}\left[\rho\left(\phi\right)\right]=\sum_{i=1}^{n}F_{Q}\left[\rho_{i}\left(\phi\right)\right]$.
For lossy NOON states, the $\rho_{D}$ of Eq.~\eqref{eq:NOON-subD}
does not contribute to the amount of QFI due to the fact that $\rho_{D}$
is independent of $\phi$. Hence the QFI for lossy NOON states is
given by $F_{Q}=F_{Q}\left[\vert\xi\left(\phi\right)\rangle\right]$
and then derived as Eq.~\eqref{eq:QFI-NOON-two} with Eq.~\eqref{eq:NOON-sub}. 

\begin{figure}
\noindent \begin{centering}
\includegraphics[scale=0.81]{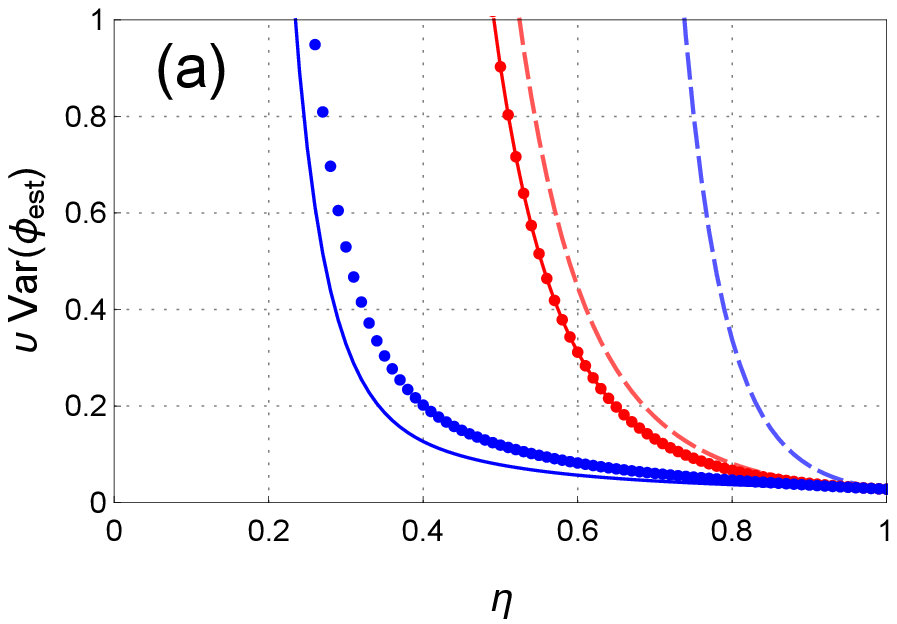}
\par\end{centering}
\noindent \begin{centering}
\includegraphics[scale=0.81]{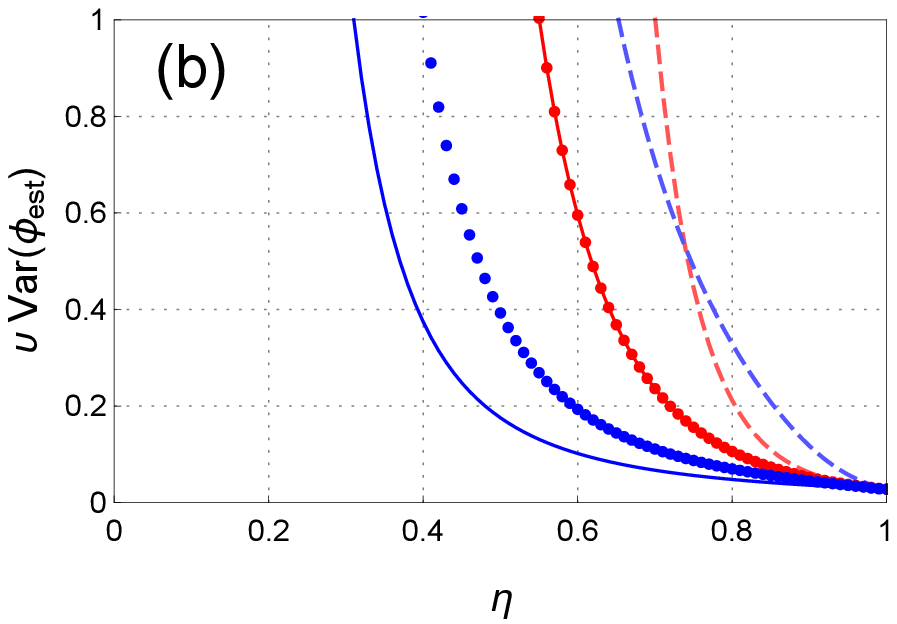}
\par\end{centering}
\centering{}\caption{(Color online) Phase sensitivities versus transmissivity $\eta$ for
losses in (a) both arms, i.e., $\eta_{a}=\eta_{b}=\eta$, and (b)
one arm, i.e., $\eta_{a}=\eta$ and $\eta_{b}=1$. Different colors
represent different probe states: Red is for $\vert6\!::\!0\rangle$
and blue is for $\vert10\!::\!4\rangle$. Solid lines correspond to
the theoretical bounds of phase sensitivity, Dashed lines refer to
the sensitivities obtained by the DP measurement and dotted lines
to those by the DPNR measurement. \label{fig:bounds}}
\end{figure}

In order to compare the performance of the two strategies in lossy
optical interferometry, we plot in Fig.~\ref{fig:bounds} the sensitivity
bounds for $\vert10\!::\!4\rangle$ and $\vert6\!::\!0\rangle$ according
to Eqs.~\eqref{eq:QFI-PEFS-two}-\eqref{eq:QFI-NOON-two}. It explicitly
indicates that $\vert10\!::\!4\rangle$ shows more robustness to photon
loss than $\vert6\!::\!0\rangle$. Although a similar conclusion was
obtained by Huver \emph{et al.} \citep{Huver2008PRA}, it was based
on a measurement scheme which was yet to be implemented. In what follows,
we address how to realize this robustness in practice with feasible
measurements.
 
\begin{figure}[t]
\begin{centering}
\includegraphics[scale=0.81]{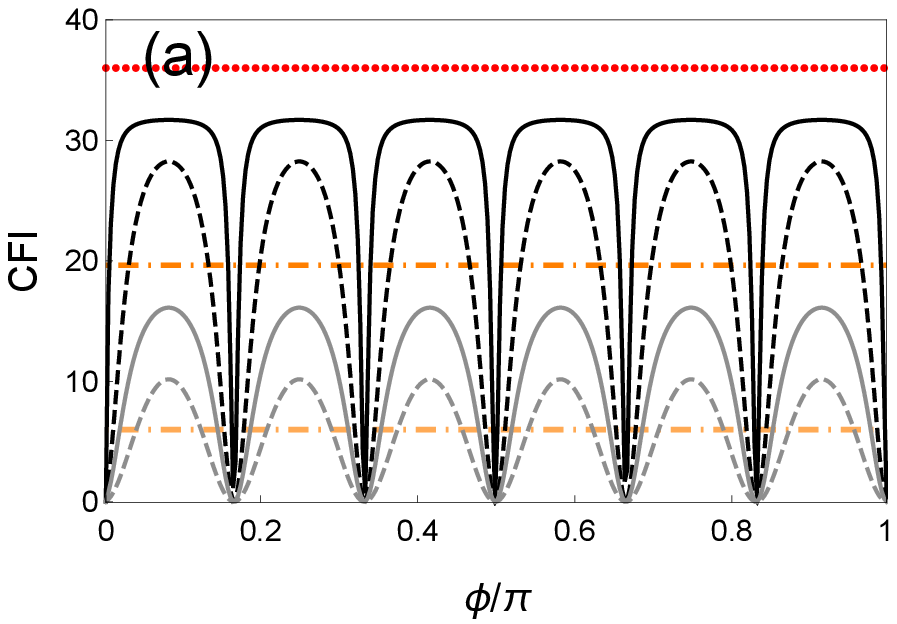}
\par\end{centering}
\begin{centering}
\includegraphics[scale=0.81]{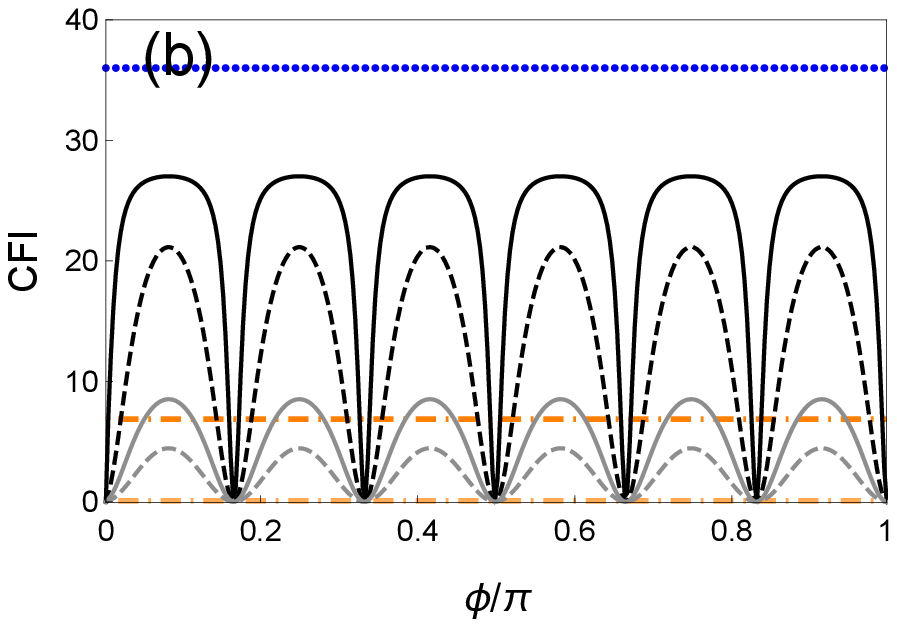}
\par\end{centering}
\centering{}\caption{(Color online) Classical Fisher information as a function of $\phi$
for (a) $\vert6\!::\!0\rangle$ and (b) $\vert10\!::\!4\rangle$ with
equal photon loss rates on each arm $\eta_{a}=\eta_{b}=\eta$. Dotted
lines represent the lossless case, i.e., $\eta=1$. The upper darker
lines for different colors and the lower lighter ones refer to the
loss cases with $\eta=0.98$ and $0.9$, respectively. Solid, dashed
and dash-dotted curves correspond to the DP, SP, and DH measurements,
respectively. \label{fig:double-parity}}
\end{figure}

\section{Phase sensitivities with three feasible measurements}

In quantum theory, a generic measurement can be described by a positive-operator-valued
measurement $\bm{\hat{M}}\equiv\{\hat{M}_{\chi}\}$ with $\chi$ the
results of the measurement. Given an operator $\bm{\hat{M}}$, the
accessible phase sensitivity is limited by a classical analog of the
inequality \eqref{eq:QCRB} $(\delta\hat{\phi})^{2}\geq(\upsilon F_{C})^{-1}$,
where $F_{C}$ is the classical Fisher information (CFI) defined as
\begin{equation}
F_{C}=\;\;\mathclap{{\displaystyle \int}}\mathclap{{\textstyle \sum}}\;\;\frac{1}{p(\chi\vert\phi)}\bigg(\frac{dp(\chi\vert\phi)}{d\phi}\bigg)^{2}d\chi,\label{eq:CFI}
\end{equation}
Here $p(\chi\vert\phi)\equiv{\rm Tr}[\hat{M}_{\chi}\rho\left(\phi\right)]$
represents the probability of the outcome $\chi$ with the specific
value of $\phi$ and the sum or integral needs to be evaluated for
discrete or continuous variables $\chi$. This bound is achievable
for the maximum-likelihood estimator with the Bayesian estimation
method when $\upsilon$ is sufficiently large \citep{Helstrom1976Book,Holevo1982Book,Uys2007PRA,Krischek2011PRL}.
If the equality $F_{C}=F_{Q}$ holds, $\hat{\bm{M}}$ is the optimal
measurement. In lossless interferometry, it has been demonstrated
that parity measurement \citep{Seshadreesan2013PRA}, photon-number-resolving
measurement \citep{Hofmann2009PRA,Lang2013PRL,Zhong2017PRA}, and
homodyne measurement \citep{Vidrighin2014NC} are optimal for all
path-symmetric pure states in the absence of noises. Below, we apply
these three types of measurements to the above phase resolution problem
and investigate the accessible sensitivities with these measurements.
We utilize a double-port measurement strategy, in which the photons
at each output port of the interferometer are simultaneously detected.
The reason is that the double-port measurement scheme may extract
more information than the single-port one (see Appendix A for a demonstration
of this superiority).

\subsection{Parity measurement}

Parity measurement was originally proposed to probe atomic frequency
in trapped ions by Bollinger \emph{et al.} \citep{Bollinger1996PRA}
and later employed for optical interferometry by Gerry \citep{Gerry2000}.
It accounts for distinguishing the states with even and odd numbers
of photons. Specifically, the parity is assigned as the value of $+1$
when the photon number of a state is even and the value of $-1$ when
odd. A parity operator acting on the output mode $c$ can be described
by \citep{Gao2010JOS,Chiruvelli2011,Jiang2012PRA}
\begin{equation}
\hat{\Pi}_{c}=\left(-1\right)^{n_{c}}=\exp\left[i\pi\left(\hat{J}_{0}+\hat{J}_{z}\right)\right].
\end{equation}
Analogously, a parity operator acting on the output mode $d$ is defined
by 
\begin{equation}
\hat{\Pi}_{d}=\left(-1\right)^{n_{d}}=\exp\left[i\pi\left(\hat{J}_{0}-\hat{J}_{z}\right)\right].
\end{equation}
Obviously, they satisfy $\hat{\Pi}_{i}^{2}=\openone\:\left(i=c,d\right)$,
with $\openone$ the identity matrix. To facilitate our calculation
below, here we consider the operation of the second BS as a part of
the measurement. Then the parity measurement on each output port through
the BS is transformed into \citep{Gao2010JOS,Chiruvelli2011,Jiang2012PRA}
\begin{eqnarray}
\hat{\pi}_{c} & = & \hat{B}\hat{\Pi}_{c}\hat{B}^{\dagger}=\sum_{N=0}^{\infty}\sum_{k=0}^{N}\vert k,N-k\rangle\langle N-k,k\vert,\label{eq:parity-a}\\
\hat{\pi}_{d} & = & \hat{B}\hat{\Pi}_{d}\hat{B}^{\dagger}=\sum_{N=0}^{\infty}\sum_{k=0}^{N}\left(-1\right)^{N}\vert k,N-k\rangle\langle N-k,k\vert.\quad\quad\label{eq:parity-b}
\end{eqnarray}
Also, we have 
\begin{eqnarray}
\hat{\pi}_{c}\hat{\pi}_{d} & = & \hat{\Pi}_{c}\hat{\Pi}_{d}=\sum_{N=0}^{\infty}\sum_{k=0}^{N}\left(-1\right)^{N}\vert k,N-k\rangle\langle k,N-k\vert,\quad\quad\label{eq:parity-ab}
\end{eqnarray}
as a result of the commutation of $\left[\hat{J}_{0},\hat{B}\right]=0$.
From above expressions, we see that both $\hat{\pi}_{c}$ and $\hat{\pi}_{d}$
are of anti-diagonal form, but slightly different up to a term $\left(-1\right)^{N}$,
and $\hat{\pi}_{c}\hat{\pi}_{d}$ is of diagonal form. Note that the
nonzero elements of $\hat{\pi}_{c}$ are all equal to $1$ and those
for $\hat{\pi}_{d}$ and $\hat{\pi}_{c}\hat{\pi}_{d}$ are assigned
alternatively $1$ and $-1$.

The SP measurement scheme has been extensively used to probe the phase
shift of the optical interferometer \citep{Seshadreesan2013PRA,Plick2010,Gerry2010,Jiang2012PRA,Tan2014PRA,Seshadreesan2011,Joo2011PRL,Campos2003,Chiruvelli2011,Anisimov2010PRL,Huang2017,Gerry2000}.
In contrast, we apply here a double-parity (DP) measurement scheme,
in which we simultaneously perform parity detection on each output
port of the interferometer. Formally, it can be expressed as a projection
operator $\hat{\bm{M}_{{\rm P}}}=\left\{ \vert\mathtt{p}_{c},\mathtt{p}_{d}\rangle\langle\mathtt{p}_{c},\mathtt{p}_{d}\vert\right\} _{\mathtt{p}_{c},\mathtt{p}_{d}=\pm1}$,
with $\left(\mathtt{p}_{c},\mathtt{p}_{d}\right)$ the results of
the parities at two output ports. We assume the second BS differs
from the first one up to a $\pi$ phase factor. The conditional probability
with respect to $\left(\mathtt{p}_{c},\mathtt{p}_{d}\right)$ is defined
by 
\begin{eqnarray}
p\left(\mathtt{p}_{c},\mathtt{p}_{d}\vert\phi\right) & = & \langle\mathtt{p}_{c},\mathtt{p}_{d}\vert\hat{B}^{\dagger}\rho\left(\phi\right)\hat{B}\vert\mathtt{p}_{c},\mathtt{p}_{d}\rangle,\label{eq:probability-parity}
\end{eqnarray}
where we have employed the cyclic property of the trace operation.
Using Eqs.~\eqref{eq:CFI} and \eqref{eq:probability-parity} yields
the CFI with respect to the DP measurement. 

To calculate exactly the CFI, we further express the $F_{C}$ into
an alternative form of Eq.~\eqref{eq:Fc-double-parity}, as shown
in Appendix B, which depends on three expectation values $\left\langle \hat{\Pi}_{c}\right\rangle $,
$\left\langle \hat{\Pi}_{d}\right\rangle $ and $\left\langle \hat{\Pi}_{c}\hat{\Pi}_{d}\right\rangle $.
With Eqs.~\eqref{eq:rho_phi} and \eqref{eq:parity-a}-\eqref{eq:parity-ab},
we obtain 
\begin{eqnarray}
\left\langle \hat{\Pi}_{c}\right\rangle  & = & D+E\cos\Delta\phi,\\
\left\langle \hat{\Pi}_{d}\right\rangle  & = & D+E\left(-1\right)^{\Delta}\cos\Delta\phi,\\
\left\langle \hat{\Pi}_{c}\hat{\Pi}_{d}\right\rangle  & = & \frac{1}{2}\left[\!\left(1-2\eta_{a}\right)^{m}\!\left(1-2\eta_{b}\right)^{n}\!\!+\!\left(1-2\eta_{a}\right)^{n}\!\left(1-2\eta_{b}\right)^{m}\!\right],\nonumber \\
\end{eqnarray}
where 
\begin{eqnarray}
D & = & \frac{1}{2}{}_{2}F_{1}\left(-m,-n;1;\frac{\eta_{a}\eta_{b}}{\gamma_{a}\gamma_{b}}\right)\left(\gamma_{a}^{m}\gamma_{b}^{n}+\gamma_{a}^{n}\gamma_{b}^{m}\right),\\
E & = & \binom{m}{\Delta}{}_{2}F_{1}\left(-n,-n;1+\Delta;\frac{\eta_{a}\eta_{b}}{\gamma_{a}\gamma_{b}}\right)\left(\eta_{a}^{\Delta/2}\eta_{b}^{\Delta/2}\gamma_{a}^{n}\gamma_{b}^{n}\right),\quad\quad
\end{eqnarray}
with $\gamma_{i}\equiv1-\eta_{i}\:\left(i=a,b\right)$ and $_{2}F_{1}\left(a,b;c;z\right)$
the ordinary hyper-geometric function \citep{Jiang2012PRA}. If photon
losses only occur on mode $a$, the above terms $D$ and $E$ can
be simplified as 

\begin{eqnarray}
D=\frac{1}{2}\binom{m}{n}\eta_{a}^{n}\gamma_{a}^{\Delta}, & \quad & E=\eta_{a}^{\Delta/2+n}
\end{eqnarray}
and $\left\langle \hat{\Pi}_{a}\hat{\Pi}_{b}\right\rangle $ becomes
\begin{eqnarray}
\left\langle \hat{\Pi}_{c}\hat{\Pi}_{d}\right\rangle  & = & \frac{\left(-1\right)^{\Delta}}{2}\left[\left(2\eta_{a}-1\right)^{m}+\left(2\eta_{a}-1\right)^{n}\right].
\end{eqnarray}

We plot in Fig.~\ref{fig:double-parity} the CFI for the single-
and double-parity measurements as a function of $\phi$ for $\vert6\!::\!0\rangle$
and $\vert10\!::\!4\rangle$. In the ideal case, i.e., $\eta=1$,
the CFI for both measurements merges and is equal to the QFI over
the whole phase interval \citep{Seshadreesan2013PRA}. In the lossy
cases, the CFI is dependent on $\phi$, in the manner of an oscillation
with a period of $\pi/\Delta$. A remarkable finding is that the DP
measurement is slightly superior to the SP measurement in these cases.
Similar results also take place in the case of losses occurring in
one arm. The difference between the two measurements can be reasonably
explained as follows. In the lossless cases, the probe states of consideration
are pure states of definite photon number. This leads to a metrological
equivalence between the SP and DP measurements. The reason is that
the value of the parity on one output mode completely depends on the
parity value from the other output mode, and thus adding one more
parity measurement leads to no additional benefit to the SP measurement.
However, the situation becomes different when losses occur. In the
lossy cases, the probe states become mixed as Eq.~\eqref{eq:rho_phi}
featuring fluctuation of the photon number. Then the results of parity
measurements on the two output ports are independent such that by
using the DP measurement can acquire more information comparing against
the use of the SP measurement. In other words, the benefit of the
DP comes from the fluctuation of the photon number (see Appendix C
for a detailed comparison between the two-parity measurement schemes).

From Figs.~\ref{fig:double-parity}(a) and \ref{fig:double-parity}(b)
we see that the decreasing rate of the CFI for $\vert10\!::\!4\rangle$
is faster than that for $\vert6\!::\!0\rangle$. This can be seen
more clearly in Fig.~\ref{fig:bounds}, where the phase sensitivities
with the DP measurement are plotted as a function of $\eta$ by setting
$\phi=\pi/2\Delta$. Such a faster deterioration accounts for the
disadvantage of PEFSs over NOON states in the presence of photon losses,
as also shown in Ref.~\citep{Jiang2012PRA}. Hence, although an additional
enhancement of sensitivity may be achievable by the DP measurement
compared with the single one, such an enhancement fails to reflect
the robustness of the PEFS strategy compared with the NOON state one. 

\subsection{Photon-number-resolving measurement}

A Photon-number-resolving (DPNR) measurement can be represented as
the projection operator $\hat{\bm{M}}_{{\rm N}}=\left\{ \vert n_{c},n_{d}\rangle\langle n_{c},n_{d}\vert\right\} _{n_{c},n_{d}=0}^{\infty}$
\citep{Pezze2008PRL,Pezze2013PRL,Lang2013PRL,Zhong2017PRA}, where
the pairs of outcomes $\left(n_{c},n_{d}\right)$ are the photon numbers
detected at the $c$ and $d$ output ports of the interferometer.
It has been shown that such a measurement is globally optimal to saturate
the quantum Cram\'er-Rao bound for all path-symmetric pure states
in the absence of photon losses \citep{Hofmann2009PRA,Zhong2017PRA}.
To evaluate the CFI with respect to the DPNR measurement, we should
determine the conditional probability in terms of $\left(n_{c},n_{d}\right)$
as

\begin{eqnarray}
p\left(n_{c},n_{d}\vert\phi\right) & = & \langle n_{c},n_{d}\vert B\rho\left(\phi\right)B^{\dagger}\vert n_{c},n_{d}\rangle.\label{eq:probability-resolving}
\end{eqnarray}
The analytical evaluation of the CFI for PEFSs is computationally
involved (except for NOON states). We thus employ numerical calculations
for the CFI. 

As mentioned in Sec. II, the lossy NOON state is given by Eq.~\eqref{eq:lossy-NOON}.
By identifying $2j=n_{c}+n_{d}$ and $2\mu=n_{c}-n_{d}$, we have
$p\left(n_{c},n_{d}\vert\phi\right)=p\left(j,\mu\vert\phi\right)$,
and then obtain  

\begin{widetext}
\begin{eqnarray}
p\left(j,\mu\vert\phi\right) & = & \sqrt{\eta_{a}^{N}\eta_{b}^{N}}\left[d_{\mu,\frac{N}{2}}^{j}\left(\frac{\pi}{2}\right)\right]^{2}\left(-1\right)^{j+\mu}\cos\left(N\phi\right)\delta_{j,\frac{N}{2}}+\frac{1}{2}\sum_{k=0}^{N}\left[d_{\mu,\frac{N-k}{2}}^{j}\left(\frac{\pi}{2}\right)\right]^{2}\binom{N}{k}\left(\eta_{a}^{N-k}\gamma_{a}^{k}+\eta_{b}^{N-k}\gamma_{b}^{k}\right)\delta_{j,\frac{N-k}{2}},\quad\quad\label{eq:probability-resolving-jm}
\end{eqnarray}
\end{widetext}where $d_{\mu,\nu}^{j}\left(\beta\right)=\langle j,\mu\vert B\vert j,\nu\rangle$
refers to the Wigner rotation function and $\delta_{i,j}$ denotes
the Kronecker delta function. Substituting the result of Eq.~\eqref{eq:probability-resolving-jm}
into Eq.~\eqref{eq:CFI} and then doing the cancellation of terms
in terms of $j\neq N/2$ as a consequence of the property of $\delta_{i,j}$,
finally yields
\begin{eqnarray}
F_{C}^{{\rm NOON}} & = & g\left(\eta_{a},\eta_{b}\right)F_{Q}^{{\rm NOON}},\label{eq:CFI-NOON}
\end{eqnarray}
with 
\begin{equation}
g\left(\eta_{a},\eta_{b}\right)=\frac{\left(\eta_{a}^{N}+\eta_{b}^{N}\right)\sin^{2}\left(N\phi\right)}{\left(\eta_{a}^{N}+\eta_{b}^{N}\right)-4\frac{\eta_{a}^{N}\eta_{b}^{N}}{\eta_{a}^{N}+\eta_{b}^{N}}\cos^{2}\left(N\phi\right)}.\label{eq:NOON-gfactor}
\end{equation}
In the above derivation, we used the identity \citep{Zhong2017PRA}
\begin{eqnarray}
\sum_{k={\rm even}}\left[d_{k-j,\frac{N}{2}}^{\frac{N}{2}}\left(\frac{\pi}{2}\right)\right]^{2} & = & \sum_{k={\rm odd}}\left[d_{k-j,\frac{N}{2}}^{\frac{N}{2}}\left(\frac{\pi}{2}\right)\right]^{2}=\frac{1}{2}.\quad\quad
\end{eqnarray}
Equation~\eqref{eq:CFI-NOON} suggests that the DPNR measurement
serves as the optimal measurement when $g\left(\eta_{a},\eta_{b}\right)=1$.
This equality holds, according to Eq.~\eqref{eq:NOON-gfactor}, when
either of the following two conditions is satisfied: (a) $\eta_{a}=\eta_{b}$;
or (b) $\phi\rightarrow\pi/2N$. In particular, for the condition
$\eta_{a}=\eta_{b}$, the equality of $F_{C}^{{\rm NOON}}=F_{Q}^{{\rm NOON}}$
always holds independently of $\phi$. This means that the DPNR measurement
is globally optimal over the whole range of values of phase parameter
for equal loss rates of two arms. Additionally, as for PEFSs, we numerically
evaluate the maximum sensitivity over $\phi$ with the DPNR measurement
in the presence of photon losses.

As shown in Fig.~\eqref{fig:bounds}, we numerically plot the phase
sensitivities achievable with the DPNR measurement for $\vert6\!::\!0\rangle$
and $\vert10\!::\!4\rangle$, respectively. Similar to the parity
measurements, the photon-number-resolving measurement also saturates
the sensitivity bounds for the two probe states in the lossless case,
i.e., $\eta=1$. This is an expected result, as demonstrated in previous
works \citep{Hofmann2009PRA,Lang2013PRL,Zhong2017PRA}. Our numerical
evidence shows that the sensitivity bounds for $\vert6\!::\!0\rangle$
is indeed reachable with the DPNR measurement for both single- and
two-arm losses, as analytically predicted above. From Fig.~\eqref{fig:bounds},
although the DPNR measurement fails to be optimal for $\vert10\!::\!4\rangle$,
it is more effective than the parity measurement; in particular, in
the case of losses on one arm, it is nearly optimal for $\vert10\!::\!4\rangle$.
More importantly, by employing the DPNR measurement, one can observe
the loss resilience of the PEFS scheme over the NOON state one in
lossy interferometry.

\subsection{Homodyne measurement}

A double-homodyne (DH) measurement can be represented as the projection
operator $\hat{\bm{M}_{{\rm H}}}=\vert x,p\rangle\langle x,p\vert$
with a pair of outcomes $\left(x,p\right)$ corresponding to the eigenvalues
of quadrature operators $\hat{X}$ and $\hat{P}$. Such a measurement
scheme was initially applied to quantum teleportation of continuous
variables \citep{Furusawa1998Science,Braunstein1998PRL}. Subsequently,
Vidrighin \emph{et al}. found that this measurement also serves as
the optimal measurement in the interferometric phase measurement with
path-symmetric pure states in the absence of photon losses \citep{Vidrighin2014NC}.
Our work also partially complements this literature, discussing the
measurement in lossy cases. 

To view its performance in our case, we need to calculate the CFI
with respect to $\hat{\bm{M}_{{\rm H}}}$. Because the results $x$
and $p$ of $\hat{\bm{M}_{{\rm H}}}$ are continuous variables, the
CFI of Eq.~\eqref{eq:CFI} should be expressed in the integral form,
\begin{eqnarray}
F_{C} & = & \int_{-\infty}^{\infty}\!\!dx\!\!\int_{-\infty}^{\infty}\!\!dp\frac{1}{p(x,p\vert\phi)}\bigg(\frac{dp(x,p\vert\phi)}{d\phi}\bigg)^{2},\label{eq:F_xp}
\end{eqnarray}
where the conditional probability in terms of $\left(x,p\right)$
is given by
\begin{equation}
p\left(x,p\vert\phi\right)=\langle n_{c},n_{d}\vert B\rho\left(\phi\right)B^{\dagger}\vert n_{c},n_{d}\rangle.\label{eq:P_xp}
\end{equation}
For the sake of our calculation, we view the projector $\hat{\bm{M}_{{\rm H}}}$
before the second BS. Each basis $\vert x,p\rangle$ hence corresponds
to $\vert\upsilon\left(x,p\right)\rangle\equiv B^{\dagger}\vert x,p\rangle$.
With the help of a discrete formulation of continuous variables \citep{Enk1999PRA,Vidrighin2014NC},
we express the transformed basis $\vert\upsilon\left(x,p\right)\rangle$,
by setting $\alpha=x+ip$, as
\begin{eqnarray}
\vert\upsilon\left(x,p\right)\rangle & = & \sum_{k,l}g_{k,l}\left(\alpha\right)\vert k,l\rangle,\label{eq:Proj_xp}
\end{eqnarray}
where
\begin{eqnarray}
g_{k,l}\left(\alpha\right) & = & \begin{cases}
\frac{e^{-\left|\alpha\right|^{2}/2}}{\sqrt{\pi}}\sqrt{\frac{l!}{k!}}\alpha^{k-l}L_{l}^{\left(k-l\right)}\left(\left|\alpha\right|^{2}\right), & k\geq l,\\
\frac{e^{-\left|\alpha\right|^{2}/2}}{\sqrt{\pi}}\sqrt{\frac{k!}{l!}}\left(-\alpha^{\ast}\right)^{l-k}L_{k}^{\left(l-k\right)}\left(\left|\alpha\right|^{2}\right), & k<l,
\end{cases}\quad\quad
\end{eqnarray}
with $L_{n}^{\left(\Delta\right)}\left(x\right)$ denoting the generalized
Laguerre polynomials. By alternatively expressing $\alpha=re^{i\varphi}$,
the above expression can be rewritten by
\begin{eqnarray}
g_{m,n}\left(r,\varphi\right) & = & g_{m,n}\left(r\right)e^{i\left(m-n\right)\varphi},
\end{eqnarray}
with
\begin{eqnarray}
g_{k,l}\left(r\right) & = & \begin{cases}
\frac{e^{-r^{2}/2}}{\sqrt{\pi}}\sqrt{\frac{l!}{k!}}r^{k-l}L_{l}^{\left(k-l\right)}\left(r^{2}\right), & m\geq n\\
\frac{e^{-r^{2}/2}}{\sqrt{\pi}}\sqrt{\frac{k!}{l!}}\left(-r\right)^{l-k}L_{k}^{\left(l-k\right)}\left(r^{2}\right), & m<n
\end{cases}\quad\quad
\end{eqnarray}
satisfying the relation $g_{l,k}\left(r\right)=g_{k,l}\left(r\right)\left(-1\right)^{k-l}$.
With Eq.~\eqref{eq:rho_phi} we obtain the conditional probability
in terms of $r$ and $\varphi$ for PEFSs as 
\begin{eqnarray}
p\left(r,\varphi\vert\phi\right) & = & \sum_{k=0}^{m}\sum_{l=0}^{n}\left(d_{1}+d_{2}\right)g_{k,l}^{2}\left(r\right)+2\cos\left[\Delta\left(\phi-2\varphi\right)\right]\nonumber \\
 &  & \times\sum_{r=0}^{n}\sum_{r^{\prime}=0}^{n}d_{3}g_{r,\Delta+r^{\prime}}\left(r\right)g_{r^{\prime},\Delta+r}\left(r\right)\left(-1\right)^{r-r^{\prime}-\Delta}.\quad\quad\label{eq:P_rphi}
\end{eqnarray}
Transforming from Cartesian coordinates $\left(x,p\right)$ to polar
coordinates $\left(r,\varphi\right)$, the CFI of Eq.~\eqref{eq:F_xp}
becomes
\begin{eqnarray}
F_{C} & = & \int_{0}^{\infty}\!\!rdr\!\!\int_{0}^{2\pi}\!\!d\varphi\frac{1}{p(r,\varphi\vert\phi)}\bigg(\frac{dp(r,\varphi\vert\phi)}{d\phi}\bigg)^{2}.\label{eq:F_rphase}
\end{eqnarray}
Submitting Eq.~\eqref{eq:P_rphi} into Eq.~\eqref{eq:F_rphase}
finally yields the CFI with respect to the DH measurement. For NOON
states, Eq.~\eqref{eq:P_rphi} can be simplified as
\begin{eqnarray}
p\left(r,\varphi\vert\phi\right) & = & \frac{1}{2}\sum_{l=0}^{N}\binom{N}{l}g_{N-l,0}^{2}\left(r\right)\left[\eta_{a}^{N}\!\!\left(\eta_{a}^{-1}\!-\!1\right)^{l}\!+\!\eta_{b}^{N}\!\!\left(\eta_{b}^{-1}\!-\!1\right)^{l}\right]\nonumber \\
 &  & +\left(-1\right)^{N}\!\!\sqrt{\eta_{a}^{N}\eta_{b}^{N}}g_{N,0}^{2}\left(r\right)\cos\left[N\left(\phi-2\varphi\right)\right].\label{eq:P_rphi_NOON}
\end{eqnarray}
For simplicity, we impose the restriction that the two arms have equal
photon loss rates (i.e., $\eta_{1}=\eta_{2}=\eta$). 

In Fig.~\ref{fig:double-parity}, we plot the phase sensitivities
achievable with the DH measurement compared against the DP measurement.
Unlike the DP measurement, the sensitivities with the DH measurement
do not vary with $\phi$. It is as clear as the fact that the value
of the integral in Eq.~\eqref{eq:F_rphase} remains unchanged by
replacing $\varphi$ with $\varphi+\phi/2$ in the way that Eqs.~\eqref{eq:P_rphi}
and \eqref{eq:P_rphi_NOON} become $\phi$-independent. What is unexpectation
is that the DH measurement exhibits poorer performance for phase sensitivity
in comparison to the DP measurement. As shown in Fig.~\ref{fig:double-parity},
a slight noise causes the sensitivities attained with the DH measurement
to decrease more significantly than that with the DP measurement.
This result is seemly contradictory to previous findings that the
homodyne detection was identified as a nearly optimal measurement
scheme in a noisy Mach-Zehnder interferometer with coherent and squeezed
vacuum light and far superior to the parity detection \citep{Gard2017EPJQT}.
A reasonable explanation for this contradictory result is that the
probe states of consideration here are non-Gaussian; in such cases
the non-Gaussian measurements (i.e., parity detection and photon-number-resolving
detection) are desirable over the Gaussian measurements (i.e., homodyne
detection). Hence, the DH measurement still fails to reflect the robustness
of the PEFS strategy compared with the NOON state one.

\section{Conclusion}

We have investigated in this paper the phase sensitivities of robust
quantum optical interferometry with PEFSs proposed by Huver \emph{et
al.} \citep{Huver2008PRA} with the help of\textcolor{teal}{{} }the
quantum estimation theory. We analytically derived the QFI for PEFSs
and NOON states in the presence of photon loss and quantitatively
demonstrated that the PEFS strategy shows more robustness to photon
loss than the NOON state strategy, which was first illustrated by
Huver \emph{et al} based on a measurement which was yet to be implemented
in experiments. 

We then addressed how to implement this advantage with three feasible
measurements, namely, parity, photon-number-resolving, and homodyne
measurement. Unlike the SP measurement scheme employed in Ref.~\citep{Jiang2012PRA},
we alternatively applied a DP measurement scheme. We found that it
can provide an additional enhancement in phase sensitivities in comparison
to the SP measurement, but it still fails to observe the robustness
of the PEFS strategy. Furthermore, we considered the DPNR measurement
and found the DPNR measurement \citep{Huelga1997PRL} to observe the
loss resilience of the PEFS strategy over the NOON state one in the
presence of loss. Interestingly, this type of measurement can always
saturate the sensitivity bounds for lossy NOON states. Finally, we
assessed the performance of the DH measurement in the aforementioned
scenario. Our results show that the DH measurement has lower performance
than the other two measurements. 

To conclude, apart from the DPNR measurement, both the DP and DH measurements
fail to indicate the robustness of the PEFSs to loss over the NOON
states. A hierarchy for the performance of the measurements of consideration
is given by $\hat{\bm{M}_{{\rm H}}}<\hat{\bm{M}_{{\rm P}}}<\hat{\bm{M}_{{\rm N}}}$
when employing the PEFSs for lossy interferometry. Our results are
readily applicable to other robust quantum metrological strategies
\citep{Dorner2009PRL,Lee2009PRA,Kolodyfmmodecutenlseniski2010PRA,Knysh2011PRA,Lee2016SR,Thekkadath2020npj},
and may shine some light on the possible detection schemes for noisy
metrology.

Finally, the primary focus of this paper is to demonstrate the robustness
of the metrological scheme with PEFSs by taking NOON states as a benchmark,
following the approach used by Huver \emph{et al} \citep{Huver2008PRA}.
An alternative way to evaluate the robustness of a metrological scheme,
as commonly used in a broad range of applications \citep{Huelga1997PRL,Escher2011NP,Chaves2012PRA,Demkowicz-Dobrzanski2012nc,Chin2012PRL,Chaves2013PRL,Froewis2014NJP,Brask2015PRX,Smirne2016PRL,Nolan2017PRL,Cai2020QIP},
is to use the separable probe state as a benchmark. Although doing
so may be beneficial to determine when both the PEFS and NOON state
schemes lose their quantum-enhanced advantage in the presence of noises,
we do not make a comparison to the separable state to stick to the
topic of the paper. Our results also shed some light on this. It was
shown in Fig. 1 that the sensitivities for both states are quite low
at a low level of loss strength, but degrade increasingly as the loss
strength increases. In other words, both states will lose their quantum
advantage as the loss strength increases. However, what is different
is the speed of degradation. Our results indicate that the sensitivity
attained by the DPNR measurement for the PEFS degrades more slowly
than that for the NOON state, which meets the theoretically predicted
performance suggested by the QFI; However, it is just the reverse
for both DP and DH measurements. These results reasonably answer the
problem raised in Introduction. 

\section*{Acknowledgments}

We are grateful to Stefan Ataman for helpful suggestions and remarks
and to the two anonymous referees for their enlightening comments
and suggestions for our paper. This work was supported by the NSFC
through Grants No. 12005106 and No. 11974189, the Natural Science
Foundation of the Jiangsu Higher Education Institutions of China under
Grant No. 20KJB140001, and a project funded by the Priority Academic
Program Development of Jiangsu Higher Education Institutions. L.Z.
acknowledges support from the China Postdoctoral Science Foundation
under Grant No. 2018M642293.

\section*{Appendix A: The superiority of the double-port measurement scheme
over the single-port measurement scheme \label{sec:AppA}}

\makeatletter 
\renewcommand{\theequation}{A\arabic{equation}} 
\makeatother 
\setcounter{equation}{0}In this appendix, we demonstrate that the double-port measurement
scheme is superior to the single-port measurement one. Here, the double-port
measurement refers to detecting the states of photons at two output
ports of the interferometer simultaneously and the single-port measurement
refers to detecting the states of photons at one of the two output
ports. Assuming that the outcome of a double-port measurement is represented
by a variable pair $\left(\chi_{c},\chi_{d}\right)$ of the detected
probability denoted by $p\left(\mathtt{\chi}_{c},\mathtt{\chi}_{d}\vert\phi\right)$,
then the marginal probability of a single outcome is determined by
\begin{equation}
p\left(\mathtt{\chi}_{c}\vert\phi\right)=\;\;\mathclap{{\displaystyle \int}}\mathclap{{\textstyle \sum}}\;\;p\left(\mathtt{\chi}_{c},\mathtt{\chi}_{d}\vert\phi\right)d\chi_{d}.\label{eq:marginal-probability}
\end{equation}
Having Eqs.~\eqref{eq:CFI} and \eqref{eq:marginal-probability},
we identify the relationship between the CFIs for the single- and
double-port measurement schemes as
\begin{eqnarray}
F_{C}^{{\rm S}} & = & \;\;\mathclap{{\displaystyle \int}}\mathclap{{\textstyle \sum}}\;\;\frac{1}{p\left(\mathtt{\chi}_{c}\vert\phi\right)}\left(\frac{dp\left(\mathtt{\chi}_{c}\vert\phi\right)}{d\phi}\right)^{2}d\chi_{c}\nonumber \\
 & = & \;\;\mathclap{{\displaystyle \int}}\mathclap{{\textstyle \sum}}\;\;\frac{1}{\;\mathclap{{\displaystyle \int}}\mathclap{{\textstyle \sum}}\;\;p\left(\mathtt{\chi}_{c},\mathtt{\chi}_{d}\vert\phi\right)d\chi_{d}}\left(\;\mathclap{{\displaystyle \int}}\mathclap{{\textstyle \sum}}\;\;\frac{dp\left(\mathtt{\chi}_{c},\mathtt{\chi}_{d}\vert\phi\right)}{d\varphi}d\chi_{d}\right)^{2}d\chi_{c}\nonumber \\
 & \leq & \;\;\mathclap{{\displaystyle \int}}\mathclap{{\textstyle \sum}}\;\;\frac{1}{p\left(\mathtt{\chi}_{c},\mathtt{\chi}_{d}\vert\phi\right)}\left(\frac{dp\left(\mathtt{\chi}_{c},\mathtt{\chi}_{d}\vert\phi\right)}{d\phi}\right)^{2}d\chi_{c}d\chi_{d}\nonumber \\
 & = & F_{C}^{{\rm D}},
\end{eqnarray}
where the above inequality comes from the use of the Cauchy-Schwarz
inequality and its equality holds if and only if 
\begin{equation}
\sqrt{p\left(\mathtt{\chi}_{c},\mathtt{\chi}_{d}\vert\phi\right)}=\frac{\lambda}{\sqrt{p\left(\mathtt{\chi}_{c},\mathtt{\chi}_{d}\vert\phi\right)}}\frac{dp\left(\mathtt{\chi}_{c},\mathtt{\chi}_{d}\vert\phi\right)}{d\varphi},
\end{equation}
is satisfied with a nonzero number $\lambda$. The above condition
is generally satisfied at some specific points of the phase or for
some typical cases (see Appendix C for an example). However, it does
not hold in most cases, in which we have $F_{C}^{{\rm D}}>F_{C}^{{\rm S}}$,
such that the double-port measurement scheme is superior to the single-port
measurement scheme in interferometric phase measurements.

\section*{Appendix B: A theoretical framework for the double parity measurement
\label{sec:AppB}}

\makeatletter 
\renewcommand{\theequation}{B\arabic{equation}} 
\makeatother 
\setcounter{equation}{0}In the following we propose a general theoretical framework for analysis
of phase sensitivity with the DP measurement. As mentioned in the
main text, the SP detection has two outcomes: $+1$ and $-1$. For
simplicity, here we denote them by $+$ and $-$, respectively. Thus,
the DP detection results in four outcomes: $++$, $+-$, $-+$, and
$--$. According to Eq.~\eqref{eq:CFI}, the CFI with the DP measurement
can be expanded to four terms as

\begin{eqnarray}
F_{C}^{{\rm D}} & = & \frac{\left[\partial_{\phi}P\left(++\vert\phi\right)\right]^{2}}{P\left(++\vert\phi\right)}+\frac{\left[\partial_{\phi}P\left(+-\vert\phi\right)\right]^{2}}{P\left(+-\vert\phi\right)}\nonumber \\
 &  & +\frac{\left[\partial_{\phi}P\left(-+\vert\phi\right)\right]^{2}}{P\left(-+\vert\phi\right)}+\frac{\left[\partial_{\phi}P\left(--\vert\phi\right)\right]^{2}}{P\left(--\vert\phi\right)}.\label{eq:CFI-DPP}
\end{eqnarray}
We find that the conditional probabilities in terms of the four outcomes
satisfy the equations

\begin{eqnarray}
P\left(+\!+\!\vert\phi\right)+P\left(+\!-\!\vert\phi\right)+P\left(-\!+\!\vert\phi\right)+P\left(-\!-\!\vert\phi\right) & = & 1,\\
P\left(+\!+\!\vert\phi\right)+P\left(+\!-\!\vert\phi\right)-P\left(-\!+\!\vert\phi\right)-P\left(-\!-\!\vert\phi\right) & = & \left\langle \Pi_{c}\right\rangle ,\\
P\left(+\!+\!\vert\phi\right)+P\left(-\!+\!\vert\phi\right)-P\left(+\!-\!\vert\phi\right)-P\left(-\!-\!\vert\phi\right) & = & \left\langle \Pi_{d}\right\rangle ,\\
P\left(+\!+\!\vert\phi\right)+P\left(-\!-\!\vert\phi\right)-P\left(+\!-\!\vert\phi\right)-P\left(-\!+\!\vert\phi\right) & = & \left\langle \Pi_{c}\Pi_{d}\right\rangle .\quad\quad
\end{eqnarray}
One can reverse these equations and solve for the conditional probabilities
in terms of the expectation values $\left\langle \Pi_{c}\right\rangle $,
$\left\langle \Pi_{d}\right\rangle $, and $\left\langle \Pi_{c}\Pi_{d}\right\rangle $.
Finally, substituting them into Eq.~\eqref{eq:CFI-DPP} yields
\begin{eqnarray}
F_{C}^{{\rm D}} & = & \frac{1}{2}\frac{\left(1+\left\langle \Pi_{c}\Pi_{d}\right\rangle \right)\left[\partial_{\phi}\left(\left\langle \Pi_{c}\right\rangle +\left\langle \Pi_{d}\right\rangle \right)\right]^{2}}{\left(1+\left\langle \Pi_{c}\Pi_{d}\right\rangle \right)^{2}-\left(\left\langle \Pi_{c}\right\rangle +\left\langle \Pi_{d}\right\rangle \right)^{2}}\nonumber \\
 &  & +\frac{1}{2}\frac{\left(1-\left\langle \Pi_{c}\Pi_{d}\right\rangle \right)\left[\partial_{\phi}\left(\left\langle \Pi_{c}\right\rangle -\left\langle \Pi_{d}\right\rangle \right)\right]^{2}}{\left(1-\left\langle \Pi_{c}\Pi_{d}\right\rangle \right)^{2}-\left(\left\langle \Pi_{c}\right\rangle -\left\langle \Pi_{d}\right\rangle \right)^{2}}.\label{eq:Fc-double-parity}
\end{eqnarray}
Defining the operators $\hat{\Pi}_{\pm}\equiv\hat{\Pi}_{c}\pm\hat{\Pi}_{d}$
and associating them with $\Pi_{\pm}^{2}=2\left(1\pm\Pi_{c}\Pi_{d}\right)$,
we can rewrite Eq.~\eqref{eq:Fc-double-parity} in a concise form
as
\begin{eqnarray}
F_{C}^{{\rm D}} & = & \sum_{i=\pm}\frac{\left\langle \hat{\Pi}_{i}^{2}\right\rangle \text{\ensuremath{\left(\partial_{\phi}\left\langle \hat{\Pi}_{i}\right\rangle \right)}}^{2}}{\frac{1}{4}\left\langle \hat{\Pi}_{i}^{2}\right\rangle ^{2}-\left\langle \hat{\Pi}_{i}\right\rangle ^{2}}.\label{eq:CFI-parity}
\end{eqnarray}
The expression \eqref{eq:Fc-double-parity} is generally divided into
two parts, based on whether the photon number of states is even or
odd. The first term in Eq.~\eqref{eq:Fc-double-parity} is contributed
by states grouped in the subspace of an even photon number and the
second term is contributed by those of an odd photon number. 

\section*{Appendix C: Comparison between the single and double parity measurements
\label{sec:AppC}}

\makeatletter 
\renewcommand{\theequation}{C\arabic{equation}} 
\makeatother 
\setcounter{equation}{0}
\begin{figure}[t]
\begin{centering}
\includegraphics[scale=0.81]{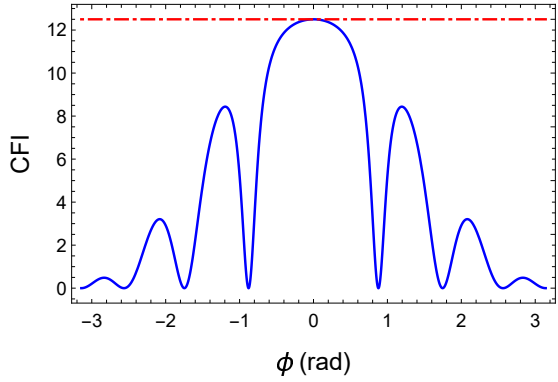}
\par\end{centering}
\centering{}\caption{(Color online) Classical Fisher information as a function of $\phi$
for $\left(\vert3\!::\!0\rangle+\vert4\!::\!0\rangle\right)/\sqrt{2}$
with the DP and SP measurements. The dash-dotted red line corresponds
to the DP measurement and the solid blue curve to the SP measurement.
\label{fig:single-vs-double-parity}}
\end{figure}
We wish to give more insight into Eq.~\eqref{eq:Fc-double-parity}.
When the total photon number of the system, denoted by $N$, is fixed,
Eq.~\eqref{eq:Fc-double-parity} can be simplified into that of the
SP measurement \citep{Seshadreesan2013PRA}

\begin{eqnarray}
F_{C}^{{\rm S}} & = & \frac{\left[\partial_{\phi}\left\langle \Pi_{c}\right\rangle \right]^{2}}{1-\left\langle \Pi_{c}\right\rangle ^{2}},\label{eq:single-parity}
\end{eqnarray}
as a result of the fact that, for even $N$, we have $\left\langle \Pi_{c}\right\rangle =\left\langle \Pi_{d}\right\rangle $
and $\left\langle \Pi_{c}\Pi_{d}\right\rangle =1$, and for odd $N$,
we have $\left\langle \Pi_{c}\right\rangle =-\left\langle \Pi_{d}\right\rangle $
and $\left\langle \Pi_{c}\Pi_{d}\right\rangle =-1$. This implies
that the DP measurement is metrologically equivalent to the SP measurement
when the number of total photons is fixed. This equivalence can be
well understood. For probe states with a fixed photon number, the
value of the parity on one output port of the interferometer is completely
determined by the parity value from the other output port; then adding
one more parity detection leads to no additional benefit. A classic
example is the NOON states, for which both the SP and DP measurements
are identified as the optimal measurement, in the manner of saturating
the Heisenberg-scaling sensitivity over the entire phase range \citep{Seshadreesan2013PRA}. 

When the fluctuation of the photon number exists, the DP and SP measurements
may not be equivalent, except at some specific phases. In order to
elucidate this, we choose a superposition between two NOON states
of even and odd photon numbers as the probe state
\begin{eqnarray}
\vert\psi\rangle & = & \sqrt{p}\vert N_{{\rm e}}\!::\!0\rangle+\sqrt{1-p}\vert N_{{\rm o}}\!::\!0\rangle.\label{eq:superposed-GHZ}
\end{eqnarray}
Obviously, this type of probe state has fluctuation of the photon
number. According to Eqs.~\eqref{eq:parity-a}-\eqref{eq:parity-ab},
we explicitly obtain the expectations with respect to Eq.~\eqref{eq:superposed-GHZ}
as
\begin{eqnarray}
\left\langle \Pi_{c}\right\rangle  & = & p\cos\left(N_{{\rm e}}\phi\right)+\left(1-p\right)\cos\left(N_{{\rm o}}\phi\right),\\
\left\langle \Pi_{d}\right\rangle  & = & p\cos\left(N_{{\rm e}}\phi\right)-\left(1-p\right)\cos\left(N_{{\rm o}}\phi\right),\\
\left\langle \Pi_{c}\Pi_{d}\right\rangle  & = & 2p-1.
\end{eqnarray}
Substituting these results into Eqs.~\eqref{eq:Fc-double-parity}
and \eqref{eq:single-parity} then gives
\begin{eqnarray}
F_{C}^{{\rm D}} & = & pN_{{\rm e}}^{2}+\left(1-p\right)N_{{\rm o}}^{2},\label{eq:FCD}\\
F_{C}^{{\rm S}} & = & \frac{\left[pN_{{\rm e}}\sin\left(N_{{\rm e}}\phi\right)+\left(1-p\right)N_{{\rm o}}\sin\left(N_{{\rm o}}\phi\right)\right]^{2}}{1-\left[p\cos\left(N_{{\rm e}}\phi\right)-\left(1-p\right)\cos\left(N_{{\rm o}}\phi\right)\right]^{2}}.\quad\label{eq:FCS}
\end{eqnarray}
Interestingly, we can find that $F_{C}^{{\rm D}}$ is identical to
$F_{Q}$ for the state of Eq.~\eqref{eq:superposed-GHZ}, meaning
that the DP measurement is an optimal measurement in the manner of
$\phi$ independence. Note that a phase-averaged operation is required
in the calculation of the QFI, since an external reference beam is
absent in the parity detection \citep{Jarzyna2012PRA}. Eqs.~\eqref{eq:FCD}
and \eqref{eq:FCS} are plotted in Fig.~\ref{fig:single-vs-double-parity}
for the case where $p=1/2$, $N_{{\rm o}}=3$, and $N_{{\rm e}}=4$,
i.e., the equal superposition of three- and four-photon NOON states.
It can been seen that the value of $F_{C}^{{\rm S}}$ is lower than
that of $F_{C}^{{\rm D}}$ over the whole phase interval except at
the zero-point location, where they have merged satisfactorily.

\bibliographystyle{apsrev4-1}
\bibliography{C:/ZW/manuscripts/me/ZW}

\end{document}